\PassOptionsToPackage{square,comma,numbers,sort&compress}{natbib}

\documentclass[final,3p,times,twocolumn]{elsarticle}
\usepackage{amssymb}
\usepackage{amsthm}

\usepackage{amsmath}
\usepackage{mathtools}

\usepackage{color}
\usepackage[colorlinks,linkcolor=blue,anchorcolor=blue,citecolor=blue,urlcolor=blue]{hyperref}

\usepackage{graphicx} %插入图片的宏包
\usepackage{float} %设置图片浮动位置的宏包
\usepackage{subfigure} %插入多图时用子图显示的宏包
\usepackage{stfloats}
\usepackage{caption}
\usepackage{booktabs}

\usepackage{orcidlink}

\begin{document}

\begin{frontmatter}



% \title{}
\title{A novel machine learning method to detect double-$\Lambda$ hypernuclear events in nuclear emulsions} 

\author[1,2]{Yan He \orcidlink{0009-0009-8519-7378}} \ead{yan.he@riken.jp}\corref{0}
% \cormark[0]

\author[2,3]{Vasyl Drozd}
\author[2]{Hiroyuki Ekawa \orcidlink{0000-0003-3507-1469}}
\author[2,4]{Samuel Escrig}
\author[2,5,6]{Yiming Gao \orcidlink{0009-0007-8595-2340}}
\author[2,7]{Ayumi Kasagi \orcidlink{0000-0003-3924-6713}}
\author[2,5,6]{Enqiang Liu}
\author[2,8]{Abdul Muneem \orcidlink{0000-0002-5316-7301}}
\author[2]{Manami Nakagawa}
\author[2,9]{Kazuma Nakazawa}
\author[4]{Christophe Rappold}
\author[2]{Nami Saito}
\author[1,2,10]{Takehiko R. Saito \orcidlink{0000-0002-7118-7349}}
\author[2,11]{Shohei Sugimoto}
\author[7]{Masato Taki}
\author[2]{Yoshiki K. Tanaka \orcidlink{0000-0002-6622-5179}}
\author[2]{He Wang}
\author[2,11]{Ayari Yanai}
\author[2,12]{Junya Yoshida}
\author[1,13]{Hongfei Zhang}

\affiliation[1]{organization={School of Nuclear Science and Technology, Lanzhou University},
                addressline={222 South Tianshui Road},
                city={Lanzhou, Gansu Province}, 
                postcode={730000},
                country={China}}
                
\affiliation[2]{organization={High Energy Nuclear Physics Laboratory, Cluster for Pioneering Research},
    addressline={RIKEN}, 
    city={Wako, Saitama},
    postcode={351-0198},
    country={Japan}}

\affiliation[3]{organization={Energy and Sustainability Research Institute Groningen, University of     
Groningen},
city={Groningen},
country={Netherlands}}

\affiliation[4]{organization={Instituto de Estructura de la Materia},
                addressline={CSIC},
                city={Madrid},
                country={Spain}}

\affiliation[5]{organization={University of Chinese Academy of Sciences},
    city={Beijing},
    postcode={100049},
    country={China}} 
    
\affiliation[6]{organization={Institute of Modern Physics, Chinese Academy of Sciences},
    addressline={509 Nanchang Road}, 
    city={Lanzhou, Gansu Province},
    postcode={730000},
    country={China}} 

\affiliation[7]{organization={Graduate School of Artificial Intelligence and Science, Rikkyo University},
                addressline={3-34-1 Nishi Ikebukuro, Toshima-ku},
                city={Tokyo},
                postcode={171-8501},
                country={Japan}}
    
\affiliation[8]{organization={Faculty of Engineering Sciences},
    addressline={Ghulam Ishaq Khan Institute of Engineering Sciences and Technology}, 
    city={Topi},
    postcode={23640},
    country={KP, Pakistan}}

% \affiliation[9]{organization={Graduate School of Engineering,Gifu University,},
%                 addressline={1-1 Yanagido},
%                 city={Gifu},
%                 postcode={501-1193},
%                 country={Japan}}

\affiliation[9]{organization={Faculty of Education, Gifu University},
                addressline={1-1 Yanagido},
                city={Gifu},
                postcode={501-1193},
                country={Japan}}
                
\affiliation[10]{organization={GSI Helmholtz Centre for Heavy Ion Research},
    addressline={Planckstrasse 1, D-64291},
    city={Darmstadt},
    country={Germany}}

\affiliation[11]{organization={Department of Physics, Saitama University},
    city={Saitama},
    postcode={338-8570}, 
    country={Japan}}

\affiliation[12]{organization={Department of physics, Tohoku University},
                address={Aramaki, Aoba-ku},
                city={Sendai},
                country={Japan}}

\affiliation[13]{organization={School of Physics, Xi’an Jiaotong University},
                city={Xi'an, shaanxi},
                country={China}}
                
\cortext[0]{Corresponding author}

\begin{abstract}
A novel method was developed to detect double-$\Lambda$ hypernuclear events in nuclear emulsions using machine learning techniques.
The object detection model, the Mask R-CNN, was trained using images generated by Monte Carlo simulations, image processing, and image-style transformation based on generative adversarial networks. 
Despite being exclusively trained on $\prescript{6\ }{\Lambda\Lambda}{\rm{He}}$ events, the model achieved a detection efficiency of 93.8$\%$ for $\prescript{6\ }{\Lambda\Lambda}{\rm{He}}$ and 82.0$\%$ for $\prescript{5\ }{\Lambda\Lambda}{\rm{H}}$ events in the produced images.
In addition, the model demonstrated its ability to detect the $\prescript{6\ }{\Lambda\Lambda}{\rm{He}}$ event named the Nagara event,
which is the only uniquely identified double-$\Lambda$ hypernuclear event reported to date. It also exhibited a proper segmentation of the event topology. 
Furthermore, after analyzing 0.2$\%$ of the entire emulsion data from the J-PARC E07 experiment utilizing the developed approach, six new candidates for double-$\Lambda$ hypernuclear events were detected, suggesting that more than 2000 double-strangeness hypernuclear events were recorded in the entire dataset. 
This method is sufficiently effective for mining more latent double-$\Lambda$ hypernuclear events recorded in nuclear emulsion sheets by reducing the time required for manual visual inspection by a factor of five hundred. 
\end{abstract}


\end{frontmatter}

% \linenumbers

\section{Introduction}
Studies on hypernuclei that contain one or more hyperons in their subatomic structure have extended our understanding of the nuclear force to the general baryon-baryon interaction under flavored SU(3) symmetry \cite{HASHIMOTO2006564, RevModPhys.88.035004}.
Hyperons, which are baryons with strange quarks, introduce a strangeness degree of freedom ($S$) into the nucleus.
A comprehensive understanding of baryon-baryon interactions involving hyperons in dense nuclear matter is crucial to elucidate the internal structure of neutron stars \cite{SCHAFFNERBIELICH2008309}.
Hypernuclear investigations are the only approach to probe baryon-baryon interactions involving hyperons in nuclear matter. 
However, experimental observations on hypernuclei remain quite limited. 
Approximately 40 single-strangeness hypernuclei ($S = -1$) have been observed. 
Particularly, experimental information on the double-strangeness ($S=-2$) sector is scarce. 
To date, only a few double-strangeness hypernuclei have been discovered \cite{10.1143/PTP.85.1287, PhysRevC.44.1905, 10.1143/ptp/85.5.951, PhysRevC.88.014003, PhysRevLett.87.212502, 10.1093/ptep/pty149, PhysRevLett.126.062501, 10.1093/ptep/ptv008, 10.1093/ptep/ptab073}. 
Among these, only the Nagara event \cite{PhysRevLett.87.212502} was uniquely identified as a double-$\Lambda$ hypernucleus, $\prescript{6\ }{\Lambda\Lambda} {\rm{He}}$ ($\Lambda\Lambda + \alpha $) in 2001 through a hybrid-emulsion experiment combining emulsion detector with spectrometers, whereas all other discovered double-$\Lambda$ hypernuclei were reported as the most likely interpretations.

Experimental studies of double-$\Lambda$ hypernuclei, where two $\Lambda$ hyperons are bound in a nucleus, are an effective approach to gain insight into the $\Lambda\Lambda$ interaction. 
The Nagara event is an epoch-making event in the study of double-$\Lambda$ hypernuclei, and provides a new and solid foundation for understanding the $\Lambda\Lambda$ interaction. 
Even today, it plays a decisive role in determining the strength of $\Lambda\Lambda$ interaction.  
Despite limited data,
the binding energy of two $\Lambda$ hyperons in the discovered double-$\Lambda$ hypernuclei appears to exhibit a linear dependence on the mass number of the double-$\Lambda$ hypernuclei \cite{Nakazawa2023}.
However, no conclusion can be drawn because of the lack of systematic studies on double-$\Lambda$ hypernuclei.
Therefore, observations of various double-$\Lambda$ hypernuclei with high accuracy are strongly awaited.
Moreover, the enhancement of the $\Lambda\Lambda$ bonding energy in double-$\Lambda$ hypernuclei owing to the three-body force represented by the $\Lambda\Lambda$-$\Xi N$ mixing effect has been highlighted by several theoretical calculations \cite{PhysRevC.69.014303,YAMAMOTO2008139}.

Nuclear emulsion experiments are one of the most efficient methods to identify double-$\Lambda$ hypernuclei by mass measurement because they make the decay chain of a double-$\Lambda$ hypernucleus visible in an emulsion with sub-$\rm{\mu}m$ spatial resolution \cite{PhysRevLett.11.26}. 
Based on the accuracy of the emulsion at the micrometer scale, it is feasible to analyze the production and sequential decays of double-$\Lambda$ hypernuclear events recorded in emulsion sheets, enabling the identification of nuclides event-by-event. 

Two events displaying a ``three-vertex" topology of sequential decays in nuclear emulsion were first reported as double-$\Lambda$ hypernuclei by Danysz et al. \cite{PhysRevLett.11.29,DANYSZ1963121} and Prowse \cite{PhysRevLett.17.782} in the 1960s.
Both events were initiated by $\Xi^-$ hyperons captured at rest by one of the nuclei in the emulsion. 
However, the $\Xi^-$ hyperon in the first event was not identified, and no photograph of the second event was reported. 
Approximately 30 years later, another double-$\Lambda$ hypernuclei event showing a clear sequential decay topology was observed in the KEK-PS E176 experiment using the hybrid emulsion method after following approximately 80 $\Xi^-$ hyperons stopped in the emulsion \cite{10.1143/ptp/85.5.951,10.1143/PTP.85.1287,PhysRevC.44.1905,AOKI2009191}. 
Although the nuclear species of this event were not uniquely identified, the existence of double-$\Lambda$ hypernuclei was first clariﬁed using the hybrid-emulsion method. 

Following the E176 experiment, the KEK-PS E373 experiment using the hybrid-emulsion method was designed to detect ten times more double-$\Lambda$ hypernuclear events than the E176 experiment. 
Finally, among nine events with sequential decay topology \cite{PhysRevC.88.014003, NAKAZAWA2010207, Nakazawa2023}, the most known event, the Nagara event \cite{PhysRevLett.87.212502} was discovered after tracking approximately 10$^3$ $\Xi^-$ hyperons stopped in the emulsion. 
From the Nagara event, the $\Lambda\Lambda$ interaction was first confirmed to be weakly attractive. 
The observation of $\prescript{6\ }{\Lambda\Lambda}{\rm{He}}$ in the ground state also imposes strict restrictions on the potential existence of H-dibaryon \cite{PhysRevLett.38.195}. 
For $\prescript{6\ }{\Lambda\Lambda}{\rm{He}}$ in which two protons and two neutrons occupy the $0s$ shell, $\Lambda\Lambda$-$\Xi N$ mixing is Pauli-suppressed.
In contrast to $\prescript{6\ }{\Lambda\Lambda}{\rm{He}}$, $\prescript{5\ }{\Lambda\Lambda}{\rm{H}}$ may have a significantly tighter $\Lambda\Lambda$ interaction strength because of $\Lambda\Lambda$-$\Xi N$ mixing \cite{PhysRevC.69.014303,YAMAMOTO2008139}. 
However, $\prescript{5\ }{\Lambda\Lambda}{\rm{H}}$ has not yet been discovered experimentally.

The J-PARC E07 experiment \cite{J-PARC_E07}, conducted recently at the Japan Proton Accelerator Research Complex (J-PARC), is the latest and most updated hybrid-emulsion experiment, and is expected to detect approximately 100 double-$\Lambda$ hypernuclei events. 
It was proposed to provide an opportunity to gather more abundant nuclear information related to strangeness as a greater variety of double-$\Lambda$ hypernuclei species. 

In the E07 experiment, $\Xi^-$ hyperons produced by the $(K^-, K^+)$ reaction were stopped and captured by the nuclei in the emulsion stacks. 
Using the hybrid-emulsion method, the position of $\Xi^-$ was tracked using other real-time detectors.
However, the detection efﬁciency of the hybrid-emulsion method for all double-strangeness hypernuclear events recorded in E07 emulsion sheets was estimated to be approximately 10$\%$ only \cite{WOS:000730546800001, Nakazawa:2013mua}. 
Owing to the limitations of spectrometer acceptance and tracking, approximately 70$\%$ of $(K^-, K^+)$ events were not tagged. 
Additionally, besides the triggered events, the `n'$(K^-, K^0)\Xi^-$ reaction, which may occur at a higher rate cannot be detected with the hybrid-emulsion method \cite{refId0}. 
Although 33 candidates for double-strangeness hypernuclear events have already been detected by following the triggered $\Xi^-$ hyperons, only three events, named Mino \cite{10.1093/ptep/pty149}, Ibuki \cite{PhysRevLett.126.062501}, and Irrawaddy \cite{10.1093/ptep/ptab073} were identified.
Therefore, it is necessary and worthwhile to develop a new detection method to achieve a significantly higher efficiency.

Approximately 1300 nuclear emulsion sheets were used in the J-PARC E07 experiment irradiated with $K^-$ beams. 
To detect all the latent double-strangeness hypernuclear events that cannot be detected using the hybrid-emulsion method, complete scanning of the entire nuclear emulsion sheets is necessary.
Recently, an overall scanning method \cite{YOSHIDA201786} that uses high-speed microscopes to capture images of an emulsion was developed. 
However, there are approximately 1.4 billion images per emulsion sheet for visual inspection, which would require over 500 years to analyze all the emulsion sheets \cite{Saito2021}.
Therefore, image recognition utilizing machine learning techniques for object detection is one of the most effective approaches for reducing the analysis time. 
Image recognition methods using machine learning techniques have already been applied to search for alpha-decay events of natural isotopes \cite{YOSHIDA2021164930} and hypertriton events \cite{KASAGI2023168663} in the emulsion sheets of the J-PARC E07 experiment successfully.
In the present study, we first applied machine learning techniques to detect double-$\Lambda$ hypernuclear events.

Section 2 describes the development procedures for both the generation of simulated double-$\Lambda$ hypernuclear events and the training of the object detection model. Section 3 describes the performance of the proposed method. Section 4 presents the results of the detection of double-$\Lambda$ hypernuclear events in E07 emulsion data. 

\begin{figure*}[htbp]
    \centering
    \includegraphics[width=0.55\textwidth]{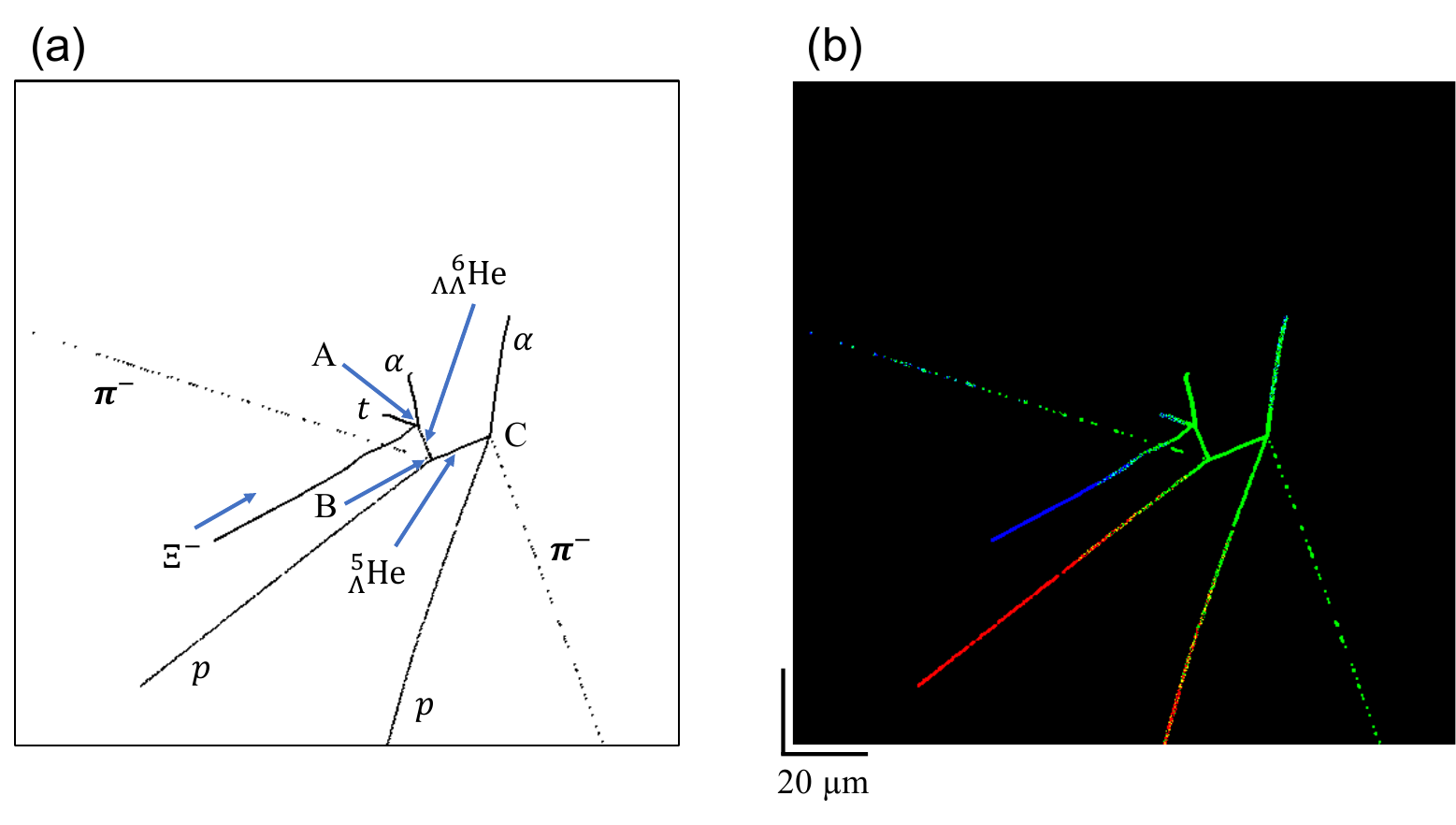}
    \caption{The images of double-$\Lambda$ hypernuclear event generated with Geant4 simulation and image processing. Panel (a) shows the trajectories and decay mode of the event. In panel (b) the trajectory information is converted to three-dimension images while RGB channels of the image represent different focus planes. The green color corresponds to the optimal focus plane, while red and blue represent the shallower and deeper plane, respectively. 
    The vertical range of the green color is 15 $\mu$m, with the center 3 $\mu$m being pure green, representing the focus plane. The remaining 6 $\mu$m on each side overlap with red (shallower plane) and blue (deeper plane), respectively, providing a gradual transition from the focused to unfocused range.}
    \label{fig1: double example}
\end{figure*}

\section{Method}
In the present work, we employed a state-of-the-art machine-learning-based object detection model, the mask region-based convolutional neural network (Mask R-CNN) \cite{he2018mask}. 
For double-$\Lambda$ hypernuclear events, there are insufficient data to train the model, as only one event has been uniquely identified to date.
% There is not real traing data   -> We have to create (done)
Therefore, we employed Geant4 Monte Carlo simulations \cite{ALLISON2016186} to generate double-$\Lambda$ hypernuclear events.
After event generation, training data containing double-$\Lambda$ hypernuclear events were produced by image processing and image style transformation, pix2pix \cite{wang2018highresolution} using generative adversarial networks (GANs) \cite{goodfellow2014generative}.
The Mask-R CNN model was trained using the produced training data.
After training, we evaluated the model, which showed sufficient efficiency in detecting double-$\Lambda$ hypernuclear events in the produced images, as discussed later in this paper.
Additionally, the model accurately detected the Nagara event and successfully segmented its topology.

\subsection{Data preparation} 
As double-$\Lambda$ hypernuclear data were insufficient to train the Mask R-CNN model, images containing double-$\Lambda$ hypernuclear events and background events were generated for training and evaluating the model by utilizing Geant4 Monte Carlo simulations, image processing, and image-style transformation. 
In the Geant4 Monte Carlo simulations, the composition of the nuclear emulsion was replicated by referring to the emulsion layer of the J-PARC E07 experiment \cite{J-PARC_E07}.
For double-$\Lambda$ hypernuclear events generated in Geant4 Monte Carlo simulations, we first considered the case of $\prescript{6\ }{\Lambda\Lambda}{\rm{He}}$, and applied the decay sequence presented in Eq. (\ref{eq1}). 
\begin{equation}
\begin{aligned}
    \Xi^- + ^{12}\rm{C} \rightarrow & \prescript{6\ }{\Lambda\Lambda}{\rm{He}} + \alpha + t \\
    &\quad \! \hookrightarrow \prescript{5\ }{\Lambda}{\rm{He}} + p + \pi^- \\
    &\qquad \quad \; \hookrightarrow \alpha + p + \pi^-
    \label{eq1}
\end{aligned}
\end{equation}
One of the $\prescript{6\ }{\Lambda\Lambda}{\rm{He}}$ events is shown in Fig. \ref{fig1: double example} (a).
As shown in the figure, $\prescript{6\ }{\Lambda\Lambda}{\rm{He}}$ is produced by $\Xi^-$ capture of $\rm^{12}C$ in the nuclear emulsion at vertex A. We assumed that $\Xi^-$ is bound in the 3D atomic orbit of $\rm^{12}C$ with a binding energy of 0.13 MeV \cite{10.1143/PTP.105.627}.
From the capture point vertex A in Fig. \ref{fig1: double example} (a), decays of $\prescript{6\ }{\Lambda\Lambda}{\rm{He}}$ and $\prescript{5}{\Lambda}{\rm{He}}$ occurred at vertices B and C, respectively. 
The decay modes of $\prescript{6\ }{\Lambda\Lambda}{\rm{He}}$ and $\prescript{5}{\Lambda}{\rm{He}}$ in Eq. (\ref{eq1}) were chosen as mesonic decay with $\pi^-$ emission, because non-mesonic decay may induce more ambiguous interpretations for identifying events. 
During the generation of an event, the lifetimes of hypernuclei in the production and decay of Eq. (\ref{eq1}) were assumed to be approximately 200 $\rm{ps}$ because the proposed method is not sensitive to the lifetime.
In addition, the mass of $\prescript{6\ }{\Lambda\Lambda}{\rm{He}}$ was defined assuming $\Lambda\Lambda$ interaction strength is zero.

When the charged particles of double-$\Lambda$ hypernuclear events undergo nuclear emulsion, the number of grains generated along the trajectory of the nuclear emulsion is correlated with their energy loss, velocity, and zenith angle \cite{10.1093/ptep/pty137}. 
In Fig. \ref{fig1: double example} (a), the thickness of the track that is related to the grain density was calculated and reproduced corresponding to the velocity and angle at each step for various tracks \cite{KASAGI2023168663}.
As the tracks in the nuclear emulsion were recorded with three dimensional information, the trajectories shown in panel (a) of Fig. \ref{fig1: double example} were converted to a colored image as shown in panel (b). RGB colors were employed to represent different focal planes, where green color indicates tracks in the optimal focus plane, while red and blue signify tracks in the shallower and deeper planes, respectively. The vertical range of the green color is 15 $\mu$m, with the center 3 $\mu$m being pure green, representing the focus plane. The remaining 6 $\mu$m on each side overlap with red (shallower plane) and blue (deeper plane), respectively, providing a gradual transition from the focused to unfocused range. The total vertical ranges of the red and blue color are 15 $\mu$m same with green color. 

\begin{figure*}[htbp]
% \begin{figure*}[ht] 
    \centering
    \includegraphics[width=0.80\textwidth]{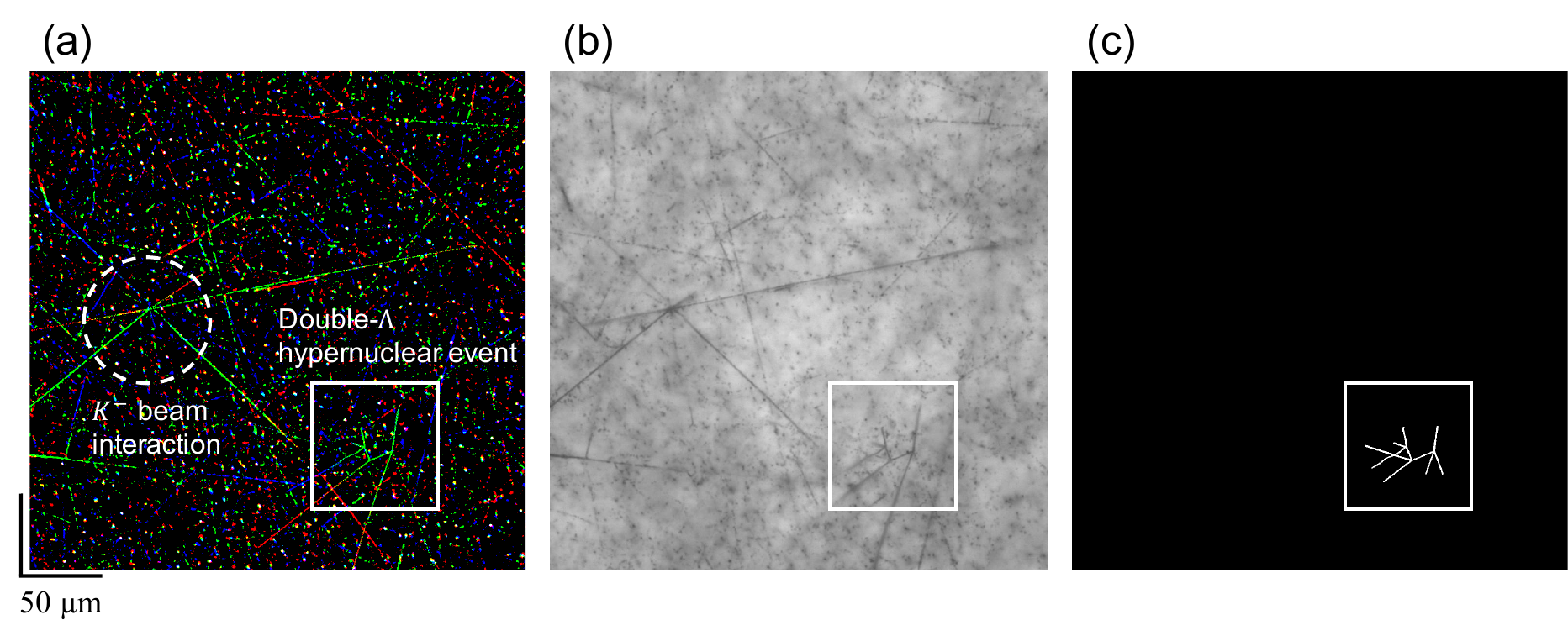}
    \caption{Panel (a) shows the colored image with RGB channels, including background and a double-$\Lambda$ hypernuclear event. Panel (b) depicts the surrogate image resembling a real emulsion image converted by the pix2pix model from the colored image, while Panel (c) shows the associated mask image, wherein only the double-$\Lambda$ hypernuclear event is marked as an object in the training data.}
    \label{Fig2: pix2pix}
\end{figure*}

In the nuclear emulsion of the E07 experiment, $K^-$ beam interaction events were the main background noise that produced tracks similar to the events of interest. 
To achieve an accurate classification and detection performance, negative samples \cite{10.1145/3394486.3403262}, $K^-$ beam interaction events, were generated as background events. 
To simulate the interaction of a $K^-$ beam at 1.8 GeV/c with the nuclides in the nuclear emulsion of the E07 experiment, the JAM package \cite{PhysRevC.61.024901}, based on data from the hadron scattering experiment was used. 
The tracks of particles from the beam interaction were visualized with the same Geant4 framework, image processing, and image-style transformation method employed in the generation of the double-$\Lambda$ hypernuclear image data.
Fig. \ref{Fig2: pix2pix} (a) shows examples of a double-$\Lambda$ hypernuclear event (marked by a solid white rectangle) and a beam interaction event (marked by a dashed circle).
Additional background tracks were extracted from the actual microscopic images of E07 emulsion data using an image filter and binarization. 
Three types of depth information for the background tracks were encoded using RGB channels, which was consistent with the method employed for the simulated images. 

After the creation of the RGB image shown in Fig. \ref{Fig2: pix2pix} (a), image style transfer using GANs \cite{goodfellow2014generative} was applied to generate emulsion images that closely mimicked real emulsion data. 
Based on the capabilities of GANs, the pix2pix model \cite{wang2018highresolution} was employed to convert the RGB image, shown in Fig. \ref{Fig2: pix2pix} (a) into an image similar to a real emulsion image as shown in Fig. \ref{Fig2: pix2pix} (b). 
The parameters for training the pix2pix model in this study are aligned with those specified in our previous work \cite{KASAGI2023168663}. 
The image produced by the trained pix2pix model in Fig. \ref{Fig2: pix2pix} (b), combined with the corresponding mask images in Fig. \ref{Fig2: pix2pix} (c), served as training data for the object detection model described in the following section.

\begin{table}[htbp]
\centering
\tabcolsep=0.2cm
\renewcommand\arraystretch{1.2}
\setlength{\abovecaptionskip}{0.2cm}
\setlength{\belowcaptionskip}{-0.2cm}
\caption{Hyperparameters for Mask R-CNN model training}
\begin{tabular}{ l l }
\hline
Parameters & Value \\  \hline %\midrule[1pt]
Backbone & ResNet50 \cite{he2015deep}  \\  
Batch size & 8 \\
Initial learning rate & 0.02  \\
Learning rate gamma & 0.1 \\ 
Learning rate step & 80, 90, 100, 110, 120  \\
momentum & 0.9 \\
Total epochs & 200 \\ \hline
\end{tabular}
\label{Table1: Mask R-CNN hyperparameters}
\end{table}

\subsection{Model training}
The proposed method employs the Mask R-CNN object detection model \cite{he2018mask}, which is a widely adopted architecture for detection and segmentation tasks \cite{rs12183015} owing to its simplicity and flexibility in network design and hyperparameter tuning. 
The model can not only detect objects of interest, but can also precisely delineate their boundaries at the pixel level, assigning confidence scores between zero and one. A score closer to one is considered to be better.

The training data for the Mask R-CNN model comprised input images paired with the corresponding mask images that labeled double-$\Lambda$ hypernuclear events as objects of interest. 
The images in Figs. \ref{Fig2: pix2pix} (b) and \ref{Fig2: pix2pix} (c) are examples of the input image and corresponding mask image, respectively. 
The mask image, outlining the shape and position of the object event can be generated using a Geant4 simulation.
A double-$\Lambda$ hypernuclear event typically displays a ``three-vertex" characterized by its sequential decay.
To ensure that the produced images maintained a clear ``three-vertex" topology for double-$\Lambda$ hypernuclear events, a cut condition of 2 $\mu$m was applied to the projected length of hypernuclear trajectories parallel to the image plane during data preparation.
In addition to the cut condition for the range of hypernuclear tracks, double-$\Lambda$ hypernuclear events with projected angles greater than 45$^\circ$ between $\prescript{6\ }{\Lambda\Lambda} {\rm{He}}$ and $\prescript{5}{\Lambda} {\rm{He}}$ were also selected. 
For the angles between the daughter particles and hypernuclei at the three vertices, a cut condition of 30$^\circ$ was applied. 
In particular, the angles between the particles and $\prescript{6\ }{\Lambda\Lambda} {\rm{He}}$ from vertex A were constrained to be greater than 30$^\circ$. 
Similarly, for the particles emitted from vertex B, the angles between the particles and both $\prescript{6\ }{\Lambda\Lambda} {\rm{He}}$ and $\prescript{5}{\Lambda} {\rm{He}}$ were also required to be greater than 30$^\circ$.
After applying these cut conditions to both the range and angles of the particles of double-$\Lambda$ hypernuclear events, a total of 18354 images were generated for training the model, with 80$\%$ allocated to the training set and the remaining 20$\%$ for validation. 

\begin{figure}[htbp]
    \centering
    \includegraphics[width=0.50\textwidth]{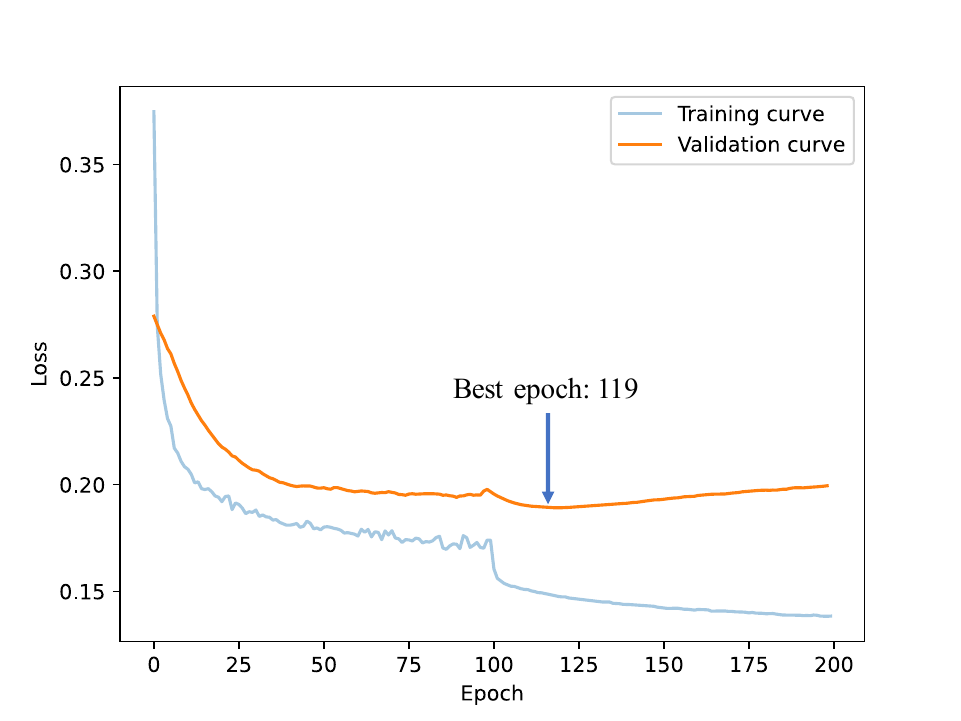}
    \caption{Plots of loss for training and validation data as functions of the number of epochs. The orange and blue lines show the loss for training and validation datasets. The epoch with the smallest validation loss value in the curve defined by Eq. (\ref{eq2}) was determined to be the best epoch, and the model at epoch = 119 was employed for inference.}
    \label{Fig3: training curve}
\end{figure}

In this study, the Mask R-CNN model was implemented using the PyTorch framework (\url{ https://github.com/multimodallearning/pytorch-mask-rcnn}). The hyperparameters applied for the model training are summarized in Table \ref{Table1: Mask R-CNN hyperparameters}. 
The smoothed validation loss \cite{YOSHIDA2021164930}
was utilized with the following recurrence formula to define the best epoch:
\begin{equation}
\begin{aligned}
    &S_0 = V_0 \\
    &S_i = wS_{i-1} + (1.0-w)V_i
\label{eq2}
\end{aligned}
\end{equation}
where, $S_i$ and $V_i$ are the $i$th smoothed and original values of the validation loss, respectively. Parameter $w$ is a weight set as 0.9, indicating the degree of smoothing. Fig. \ref{Fig3: training curve} shows the training and validation losses during model training, represented by blue and orange lines, respectively.
Epoch 119, characterized by the lowest smoothed loss, was determined to be the optimal epoch for subsequent inference.

\section{Model performance}
\begin{figure*}[htbp]
    \centering
    \includegraphics[width=0.50\textwidth]{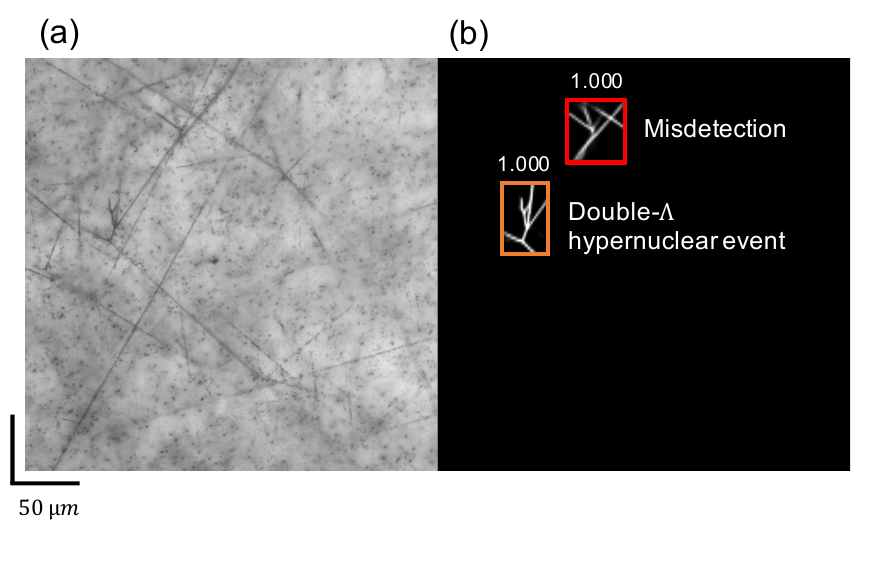}
    \caption{The example of the object detected by the model for produced images. While the model detected double-$\Lambda$ hypernuclear events, misdetection occurred during evaluation with test dataset. The misdetections are primarily caused by beam interactions that create tracks crossing with other tracks as illustrated in this figure. Panel (a) is the produced emulsion images and panel (b) shows the mask image highlighting the objects detected by the developed model. The objects in the masks image are detected double-$\Lambda$ hypernuclear event and misdetection indicated by the orange and red boxes, respectively.}
    \label{Fig4: detected_double}
\end{figure*}

After training the model, we first evaluated its performance by analyzing the produced images. 
To verify whether the model can detect double-$\Lambda$ hypernuclear events, four datasets containing 500 images each were employed for evaluation. 
These images included both double-$\Lambda$ hypernuclear and background events. These events were generated using the same procedure as that used to produce the training and validation datasets. 
The first group of images featured $\prescript{6\ }{\Lambda\Lambda}{\rm{He}}$ events with the decay mode described by Eq. (\ref{eq1}), which was identical to that of the training dataset. 
Although the model was trained exclusively on $\prescript{6\ }{\Lambda\Lambda}{\rm{He}}$ events with the decay mode shown in Eq. (\ref{eq1}), double-$\Lambda$ hypernuclear events with other decay modes described below were also used for evaluation.
The decay mode of $\prescript{6\ }{\Lambda\Lambda}{\rm{He}}$ events in the second group is as follows:
\begin{equation}
    \label{eq3}
    \begin{split}
        \Xi^- + ^{12}\rm{C} \rightarrow 
        & \prescript{6\ }{\Lambda\Lambda}{\rm{He}} + \alpha + t \\
        & \quad \! \hookrightarrow \prescript{5\,}{\Lambda}{\rm{He}} + p + \pi^- \\
        & \qquad \quad \; \hookrightarrow t + p + n
    \end{split}
\end{equation}
In addition to $\prescript{6\ }{\Lambda\Lambda}{\rm{He}}$ events, $\prescript{5\ }{\Lambda\Lambda}{\rm{He}}$ events with the following two decay modes for the remaining two groups of images were used for model evaluation.
\begin{equation}
    \label{eq4}
    \begin{aligned}
        \quad \Xi^- + ^{12}\rm{C} \rightarrow 
        & \prescript{5\ }{\Lambda\Lambda}{\rm{H}} + \alpha + \alpha \\
        & \quad \! \hookrightarrow \prescript{5\,}{\Lambda}{\rm{He}} + \pi^- \\
        & \qquad \quad \; \hookrightarrow \alpha + p + \pi^-
    \end{aligned}
\end{equation}
\begin{equation}
    \label{eq5}
    \begin{aligned}
        \qquad \Xi^- + ^{12}\rm{C} \rightarrow
        & \prescript{5\ }{\Lambda\Lambda}{\rm{H}} + \alpha + \alpha \\
        & \quad \! \hookrightarrow \prescript{4\,}{\Lambda}{\rm{He}} + n + \pi^- \\
        & \qquad \quad \; \hookrightarrow \rm{^3He} + p + \pi^-
    \end{aligned}
\end{equation}
When the four groups of images were input into the model, double-$\Lambda$ hypernuclear events were successfully detected, as indicated by the orange box in the mask image in Fig. \ref{Fig4: detected_double}.
However, the model also produced some misdetections, primarily due to beam interactions crossing other tracks, such as the object highlighted by the red box in the mask image in Fig. \ref{Fig4: detected_double}. 

\begin{figure*}[htbp]
    \centering
    \includegraphics[width=0.75\textwidth]{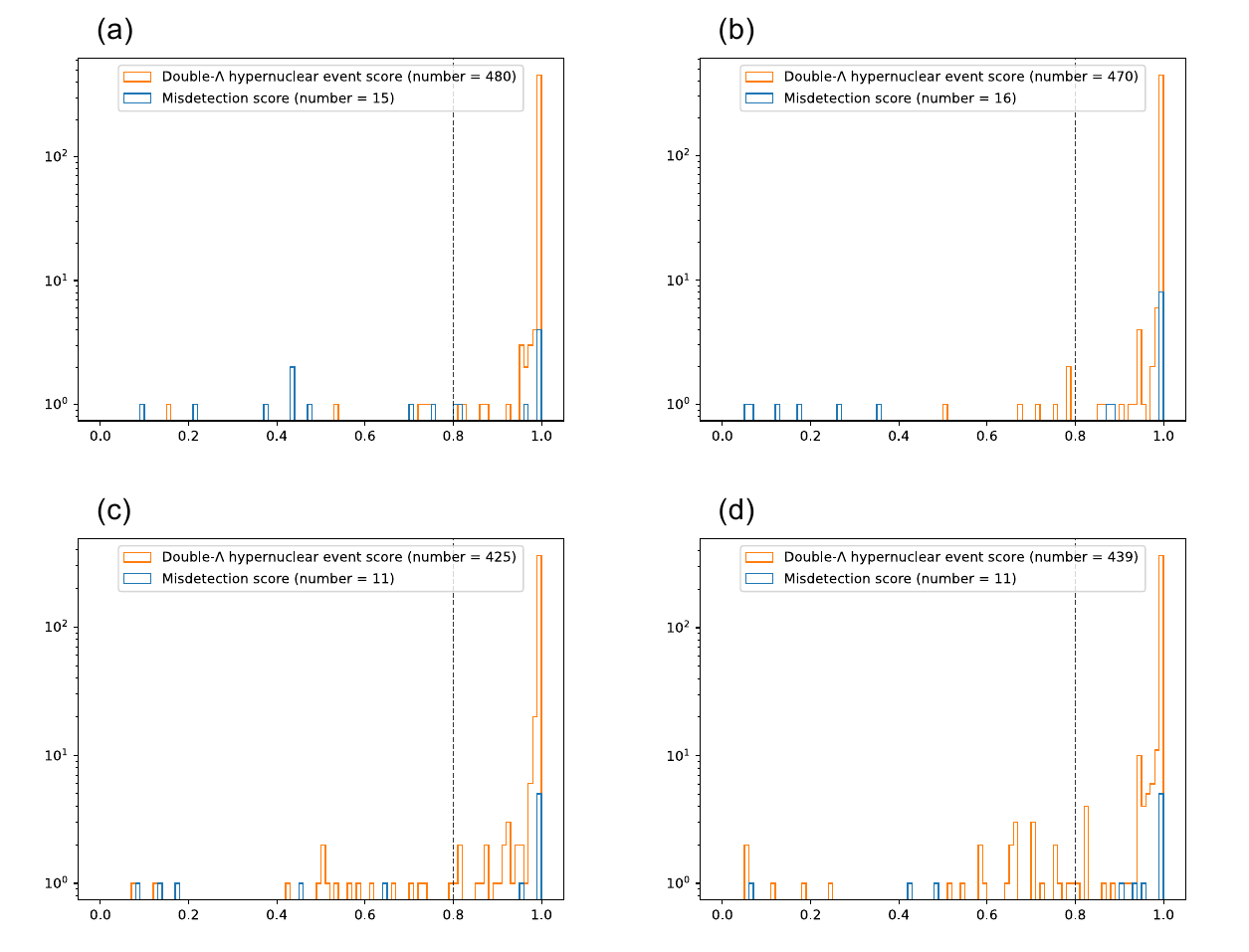}
    \caption{Score distributions of the detected double-$\Lambda$ hypernuclear events and misdetections for the four test datasets. Each dataset contains 500 images, and each image includes double-$\Lambda$ hypernuclear event with a specific decay mode. Double-$\Lambda$ hypernuclear events in panel (a) and (b) are $\prescript{6\ }{\Lambda\Lambda}{\rm{He}}$ events with decay mode Eq. (\ref{eq1}) and Eq. (\ref{eq3}), and double-$\Lambda$ hypernuclear events in (c) and (d) are $\prescript{5\ }{\Lambda\Lambda}{\rm{H}}$ with decay mode Eq. (\ref{eq4}) and Eq. (\ref{eq5}). 
    Orange lines represent the score distributions of detected double-$\Lambda$ hypernuclear events and blue lines show the score distribution of misdetections. With a score threshold of 0.8 as shown with black dash lines, the detection efficiency and purity of the model for test datasets were calculated with Eq. (\ref{eq6}) and Eq. (\ref{eq7}), respectively. The results of calculation are listed in Table. \ref{Table2: efficiency and purity}.}
    \label{Fig5: score_distribution}
\end{figure*}

\begin{figure*}[htbp]
    \centering
    \includegraphics[width=0.55\textwidth]{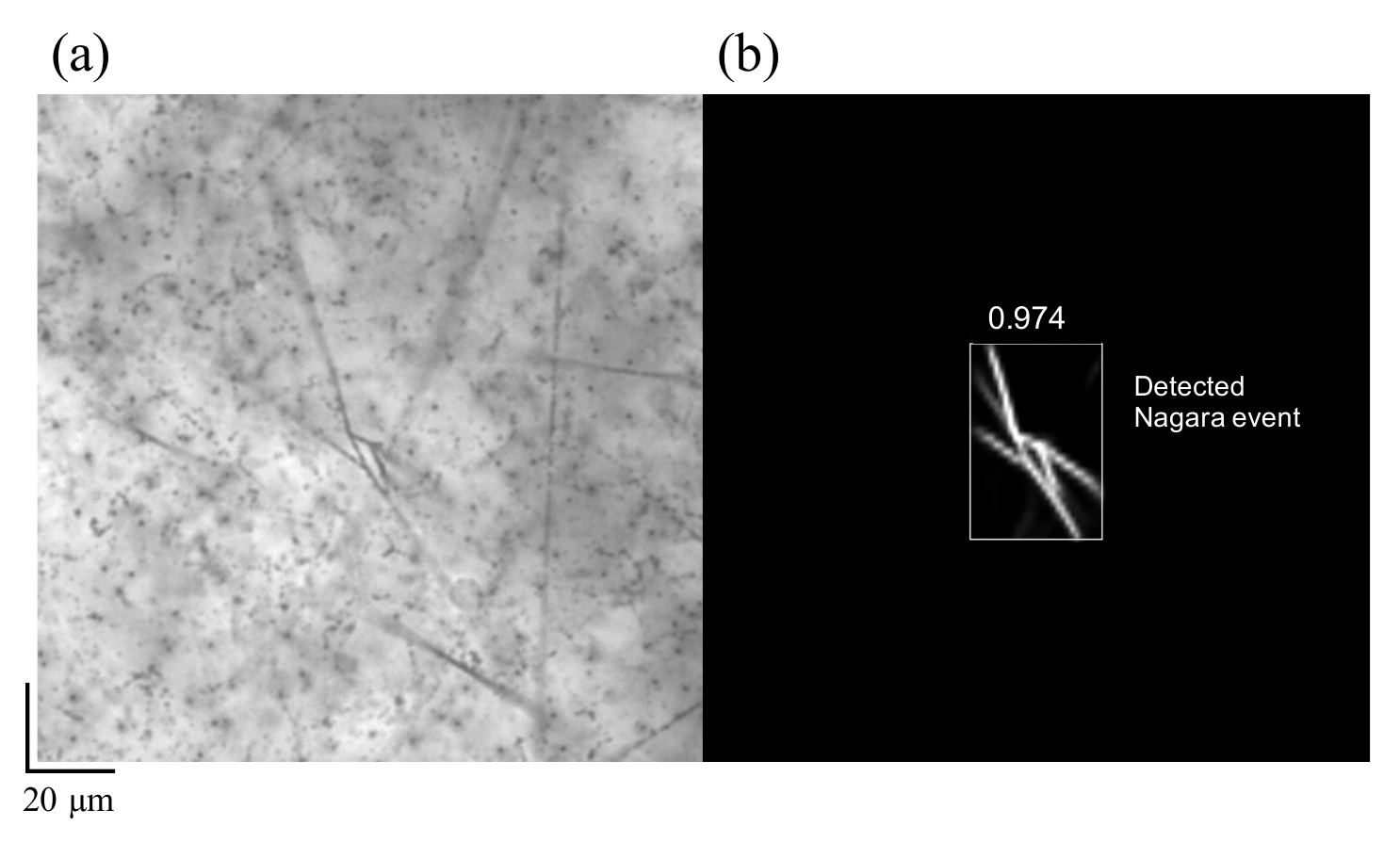}
    \caption{Detection reslut of Nagara event \cite{PhysRevLett.87.212502} by the developed model. Panel (a) shows the emulsion image captured under microscope with a 20$\times$ objective lens. Panel (b) shows the mask image detected by the model. White pixels are detected $\prescript{6\ }{\Lambda\Lambda}{\rm{He}}$ event. The mask image clearly visualizes both the production and decay vertices of the event with a score of 0.974, and all emitted particle tracks are segmented precisely.}
    \label{Fig6: Nagara_detection}
\end{figure*}

Fig. \ref{Fig5: score_distribution} shows the score distributions of the test results for the four datasets.
Each dataset consisted of 500 images, including double-$\Lambda$ hypernuclear events with a specific decay mode: Eq. (\ref{eq1}) in (a), Eq. (\ref{eq3}) in (b), Eq. (\ref{eq4}) in (c), and Eq. (\ref{eq5}) in (d). 
The blue line represents the distribution of misdetections, whereas the orange line shows the score distribution of the detected double-$\Lambda$ hypernuclear events.
Using a score threshold of 0.8, the detection efficiency and purity of the model for the test datasets were calculated as
\begin{equation}
\label{eq6}
\rm Efficiency = \frac{N_{detected\text{-}double}}{N_{total\text{-}double}}
\end{equation}
\begin{equation}
\label{eq7}
\rm Purity = \frac{N_{detected\text{-}double}}{N_{detected\text{-}double} + N_{misdetection}}
\end{equation}
$\rm{N_{detected\text{-}double}}$ represents the number of detected double-$\Lambda$ hypernuclear events, $\rm{N_{total\text{-}double}}$ is the total number of double-$\Lambda$ hypernuclear events in the test dataset, and $\rm{N_{misdetection}}$ is the number of misdetections. The efficiencies and purities of the four test datasets are presented in Table \ref{Table2: efficiency and purity}. 

For the decay modes described in Eq. (\ref{eq1}) and Eq. (\ref{eq3}), $\prescript{6\ }{\Lambda\Lambda}{\rm{He}}$ is produced by $\Xi^{-}$ capture at rest by $\rm ^{12}C$ and decays with $\pi^{-}$ emission. 
The primary distinction between the two decay modes lies in the subsequent decay of a single-$\Lambda$ hypernucleus.
\begin{itemize}
    \item in Eq. (\ref{eq1}), $\prescript{5}{\Lambda}{\rm{He}}$ decays with $\pi^-$ emission.
    \item in Eq. (\ref{eq3}), non-mesonic decay occurs for $\prescript{5}{\Lambda}{\rm{He}}$, as observed in the Nagara event \cite{PhysRevLett.87.212502}.
\end{itemize}
For $\prescript{6\ }{\Lambda\Lambda}{\rm{He}}$ events with the decay modes described by Eq. (\ref{eq1}) and Eq. (\ref{eq3}), the model achieved detection efficiencies of 94.8 $\%$ and 92.8 $\%$, respectively, along with purities of 98.5 $\%$ and 97.9 $\%$. 
On average, the model achieved a detection efficiency of 93.8$\%$ and purity of 98.2$\%$ for $\prescript{6\ }{\Lambda\Lambda}{\rm{He}}$.

In addition, $\prescript{5\ }{\Lambda\Lambda}{\rm{H}}$ events with decay modes described by Eqs. (\ref{eq4}) and (\ref{eq5}) were used to further evaluate the model performance. 
$\prescript{5\ }{\Lambda\Lambda}{\rm{H}}$ was produced by $\Xi^{-}$ capture at rest by $\rm ^{12}C$, followed by sequential mesonic decay. 
The distinction between Eq. (\ref{eq4}) and Eq. (\ref{eq5}) lies within the neutron emission during the $\prescript{5\ }{\Lambda\Lambda}{\rm{H}}$ decay.
\begin{itemize}
    \item in Eq. (\ref{eq4}), $\prescript{5\ }{\Lambda\Lambda}{\rm{H}}$ decays without neutron emission.
    \item in Eq. (\ref{eq5}), $\prescript{5\ }{\Lambda\Lambda}{\rm{H}}$ decays with neutron emission.
\end{itemize}
Although trained only on $\prescript{6\ }{\Lambda\Lambda}{\rm{He}}$ events, even for $\prescript{5\ }{\Lambda\Lambda}{\rm{H}}$, the model exhibited a high detection efficiency of 81.4$\%$ and purity of 98.5$\%$ for the decay mode Eq. (\ref{eq4}), and 82.6$\%$ efficiency and 98.1$\%$ purity for the decay mode Eq. (\ref{eq5}).
On average, the model achieved a detection efficiency of 82.0$\%$ and a purity of 98.3$\%$ for $\prescript{5\ }{\Lambda\Lambda}{\rm{H}}$.

After evaluating the model using the generated images, we tested it using the Nagara event \cite{PhysRevLett.87.212502}.
Fig. \ref{Fig6: Nagara_detection} presents the detection results. 
Panel (a) shows a microscopic image of the Nagara event in the nuclear emulsion captured with a 20$\times$ objective lens, and panel (b) displays the mask image predicted by the model, highlighting the detected $\prescript{6\ }{\Lambda\Lambda}{\rm{He}}$ event.
The model successfully detected the Nagara event with a score of 0.974. 
The corresponding mask image clearly visualizes both the production and decay vertices, accurately segmenting all the emitted particle tracks.

\begin{table*}[htbp]
\centering
\tabcolsep=0.4cm
\renewcommand\arraystretch{1.2}
\setlength{\abovecaptionskip}{0.2cm}
\setlength{\belowcaptionskip}{-0.2cm}
\caption{Detection efficiency and purity of the model for $\prescript{6\ }{\Lambda\Lambda}{\rm{He}}$ events with decay mode Eq. (\ref{eq1}) and Eq. (\ref{eq3}), and $\prescript{5\ }{\Lambda\Lambda}{\rm{H}}$ events with decay mode Eq. (\ref{eq4}) and Eq. (\ref{eq5}) with a score threshold 0.8. $\rm{N_{detected\text{-}double}}$ is the number of the double-$\Lambda$ hypernuclear events detected by the model, and $\rm{N_{misdetection}}$ is the number of the misdetections. Efficiency and purity are calculated by the Eq. \ref{eq6} and Eq. \ref{eq7}.}
\begin{tabular}{ c c c c c c}
\hline
Double-$\Lambda$ hypernucleaus & Decay mode & $\rm{N_{detected\text{-}double}}$ & $\rm{N_{misdetection}}$& Efficiency  & Purity \\  \hline %\midrule[1pt]
$\prescript{6\ }{\Lambda\Lambda}{\rm{He}}$ & Eq. (\ref{eq1}) & 474 & 7  & 94.8$\%$ & 98.5$\%$ \\  
$\prescript{6\ }{\Lambda\Lambda}{\rm{He}}$ & Eq. (\ref{eq3}) & 464 & 10 & 92.8$\%$ & 97.9$\%$ \\
$\prescript{5\ }{\Lambda\Lambda}{\rm{H}}$ & Eq. (\ref{eq4})  & 407 & 6  & 81.4$\%$ & 98.5$\%$  \\
$\prescript{5\ }{\Lambda\Lambda}{\rm{H}}$ & Eq. (\ref{eq5})  & 413 & 8  & 82.6$\%$ & 98.1$\%$ \\ \hline
\end{tabular}
\label{Table2: efficiency and purity}
\end{table*}

\section{Results and discussions}
Following the evaluation of the produced images, the model performance was evaluated on actual emulsion data. 
The evaluation employed 2.4 million microscopic images acquired from a volume of approximately $\rm{25\ cm^2 \times 0.025\ cm}$ on the emulsion sheet in the J-PARC E07 experiment. 
From these images, the model detected 8336 images that exhibited characteristics similar to the ``three-vertex" topology.

The emulsion images used as inputs for the model were captured by optical scanning.
The emulsion sheet was scanned using a microscope with a 20$\times$ objective lens, moving in the horizontal and vertical directions to capture images from different regions. 
To acquire images from different focal depths, the stage was moved perpendicular to the emulsion sheet in approximately 3 $\rm{\mu m}$ steps.
This scanning process can count an event multiple times if it appears across multiple focal planes. 
To address this issue, duplicate events were removed based on the object positions predicted by the model. 
If the distance between the positions of the objects detected in adjacent focal planes was less than 30 $\mu m$, the latter object was considered as a duplicate and was removed.  
Additionally, the model detected the dust captured in the emulsion, which was removed based on the number of black pixels in the mask images detected by the model \cite{KASAGI2023168663}. 
From the initial 8336 detected images, 3343 objects were duplicates and 1091 containing dust were excluded. 
Ultimately, 4177 objects remained in 3902 images for further visual inspection.

\begin{figure*}[htbp]
    \centering
    \includegraphics[width=0.9\textwidth]{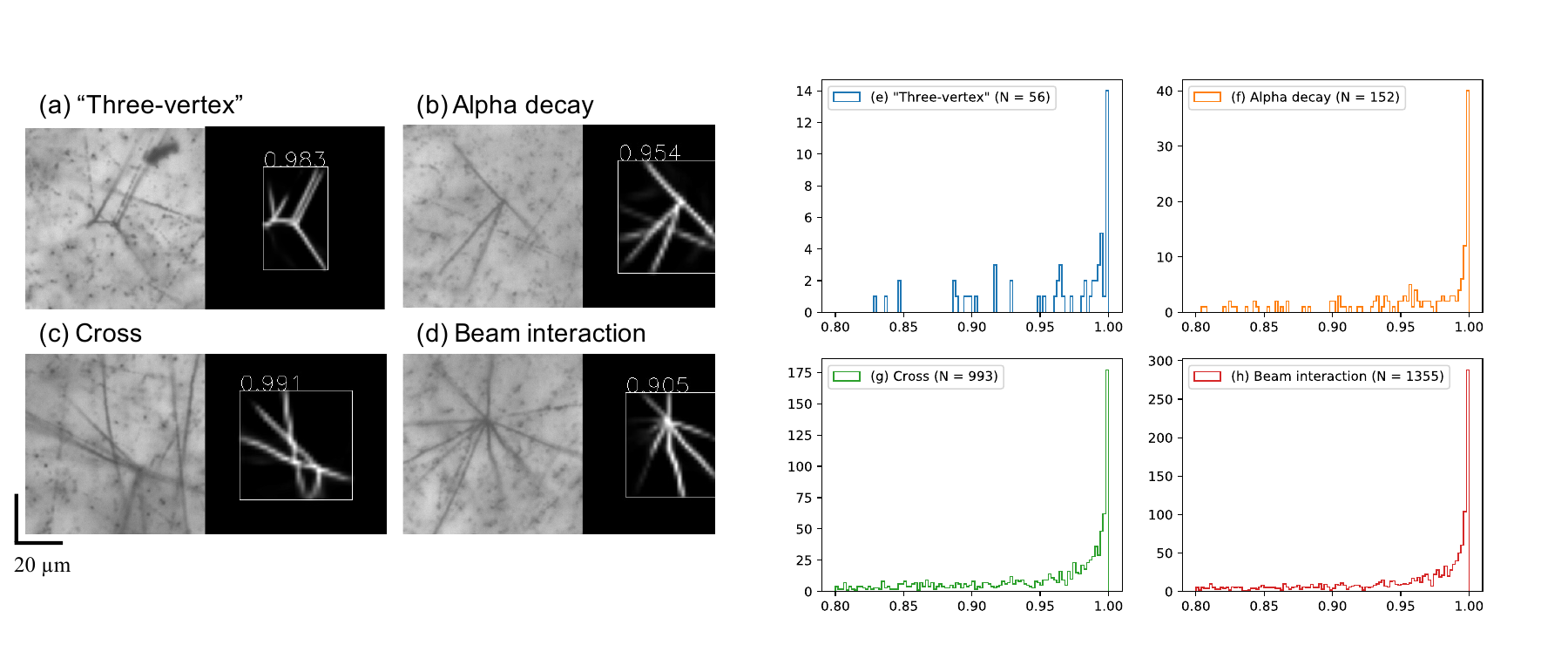}
    \caption{Examples and score distributions of the detected objects by the proposed model. Each pair of images in left part displays the emulsion images (left) and corresponding mask images (right). Panel (a) demonstrates a positive detection of the ``three-vertex" event, and panel (b), (c), (d) are examples of the alpha decay, cross and beam interaction events, respectively. The right panel presents histograms of the score distributions for the four detected objects categories in (e), (f), (g) and (h), corresponding to the categories shown in (a), (b), (c), and (d).}
    \label{Fig7: catogery_and_score}
\end{figure*}

Fig. \ref{Fig7: catogery_and_score} shows examples (left) and score distributions (right) of the four object categories detected using the developed model.
Each pair of example images displays the emulsion images captured under a 20$\times$ objective lens and mask images from the model detection.
Panel (a) shows an example of positive detection of the ``three-vertex" event. During visual inspection of the images captured using a 20$\times$ objective lens, objects that appeared to be potential ``three-vertex" events were classified for further examination under a higher-magnification lens. 
In total, there were 56 positive objects of ``three-vertex" events from 4177 detected objects. 
Beyond the ``three-vertex" events, the model detected additional events with the following categories:
\begin{itemize}
    \item 152 alpha decay events as shown in Fig. \ref{Fig7: catogery_and_score} (b);
    \item 993 objects with at least two vertices caused by cross tracks in Fig. \ref{Fig7: catogery_and_score} (c);
    \item 1355 beam interaction events in Fig. \ref{Fig7: catogery_and_score} (d).
\end{itemize}
The right panel of Fig. \ref{Fig7: catogery_and_score} displays the score distribution of these four categories of objects on a logarithmic scale. The remaining objects detected were dust and duplicates. 

The developed model reduced the number of background images from 2.4 million to 4177, which were retained for visual inspection, representing a reduction factor of $1.7\times10^{-3}$. 
This time consumption is 500 times shorter than the 500 years required for manual visual inspection of the entire nuclear emulsion, as discussed in Section 1.
Consequently, it is feasible to visually inspect all images of the entire J-PARC E07 nuclear emulsion within one year.

After the evaluation, we applied the model to approximately 0.2$\%$ of the entire E07 emulsion dataset captured from a volume of $\rm{4800 \ cm^2 \ \times \ 0.025 \ cm}$ of the emulsion sheets. In total, 12962 potential ``three-vertex" objects were classified. From these objects, six double-$\Lambda$ hypernuclear candidates were observed after reviewing with a 90$\times$ objective lens. The remaining classified ``three-vertex" objects primarily consisted of beam interactions that intersected with other tracks, creating a ``three-vertex" appearance that was difficult to distinguish from the images captured using a 20$\times$ objective lens. 

\begin{figure*}[htbp]
    \centering
    \includegraphics[width=0.65\textwidth]{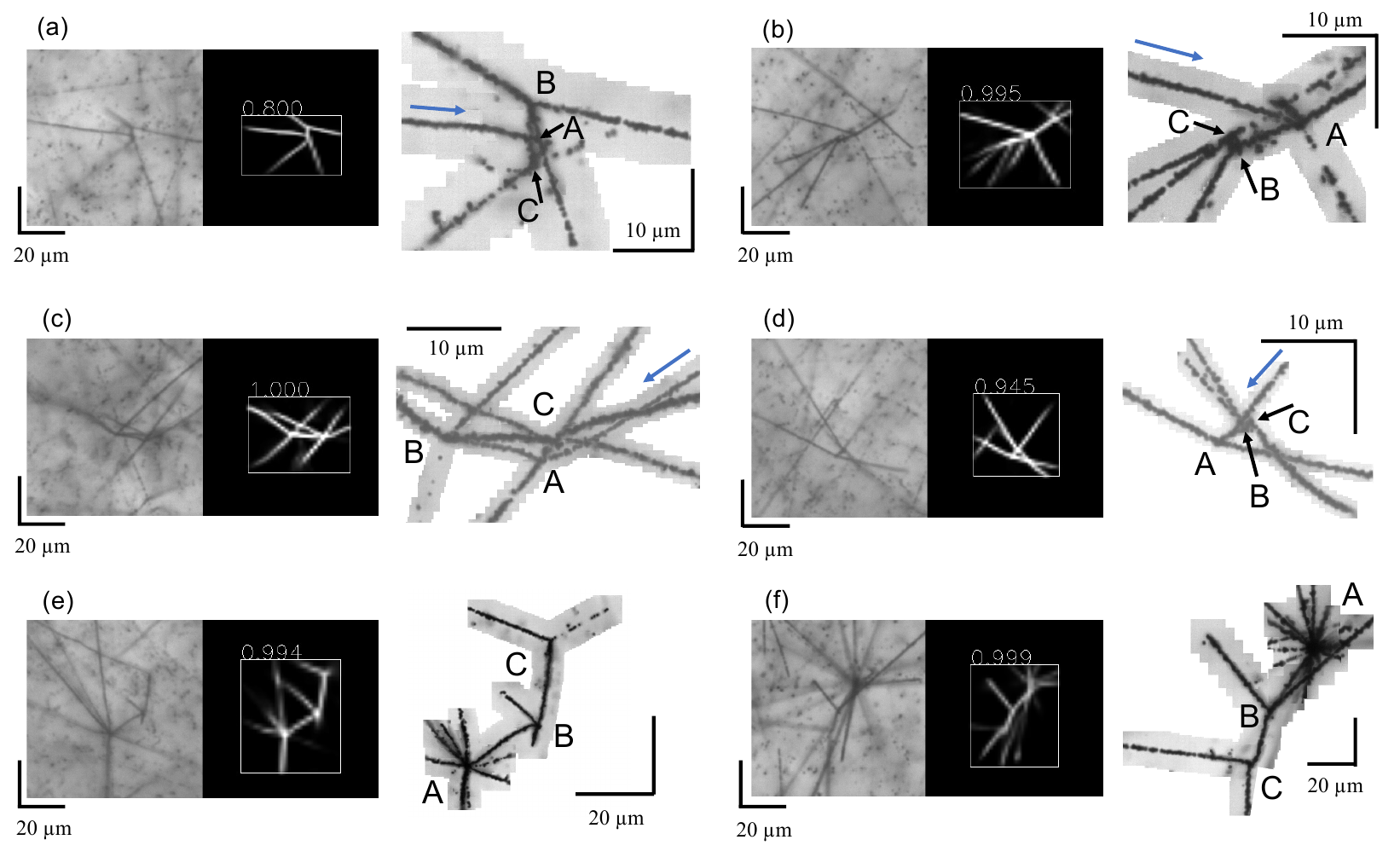}
    \caption{The images of six candidates detected by the developed model. Panel (a-f) presents six groups of images representing six candidates. The left-hand images in each group, used as model input, were captured under a 20$\times$ objective lens. The middle images are mask images detected by the developed model. The rightmost images present the event topology captured with a 90$\times$ objective lens. The blue arrows in the right images of panels (a), (b), (c) and (d) show the incoming particle of the events. And the incoming particles were captured at the first vertex A. After capture, the sequential decay occurred at vertex B and C, respectively. In panels (e) and (f), vertex A represents a beam interaction followed by cascade decays at vertex B and C. }
    \label{Fig8: candidates}
\end{figure*}

Fig. \ref{Fig8: candidates} (a-f) shows images of the six candidates.
The left panels of the images in each group show the detection results of the developed model.
The rightmost images depict the event topology under a microscope with a 90$\times$ objective lens. 
The blue arrows in the rightmost images in panels (a), (b), (c), and (d) indicate the incoming particles of the events.
The incoming particles were captured at the first vertex A, and sequential decay occurred at vertices B and C.
In panels (e) and (f), vertex A represents the beam interaction, followed by cascade decay at vertices B and C. 
These six candidates of double-strangeness hypernuclear events showed clear ``three-vertex" topology and cascade decays occurred. 
Based on the number of detected candidates, it is estimated that the total dataset includes over 2000 double-strangeness hypernuclear events.
Further kinematic analyses are required to definitively identify these events.
Detailed analyses of these events are currently underway.

\section{Summary}
In this study, we developed a novel method utilizing Geant4 Monte Carlo simulations, image processing, and image-style transformation with GANs to detect double-$\Lambda$ hypernuclear events in nuclear emulsion sheets of the J-PARC E07 experiment.
The proposed method can detect double-$\Lambda$ hypernuclear events in both produced and actual emulsion images. 
For the produced images, the method achieved detection efficiencies of 93.8$\%$ and 82.0$\%$ for $\prescript{6\ }{\Lambda\Lambda}{\rm{He}}$ and $\prescript{5\ }{\Lambda\Lambda}{\rm{H}}$, respectively, with corresponding purities of 98.2$\%$ and 98.3$\%$. 
In addition to the produced images, the proposed method successfully detected the Nagara event with a confidence score of 0.974.
When applied to E07 emulsion images, the method drastically reduced the background images by a factor of 0.0017 and successfully detected six candidates of double-$\Lambda$ hypernuclear events over 0.2$\%$ of the full nuclear emulsion dataset of the E07 experiment. 
The number of detected candidates suggests that more than 2000 double-strangeness hypernuclear events were recorded in the entire dataset. 
The proposed method shows considerable promise for application across the entire E07 nuclear emulsion dataset, potentially enhancing visual inspection efficiency by approximately 500 times.


\section*{CRediT authorship contribution statement}
\textbf{Yan He:} Conceptualization, Methodology, Software, Validation, Formal analysis, Investigation, Data curation, Writing – original draft, Writing – review \& editing, Visualization. 
\textbf{Vasyl Drozd:} Methodology, Formal analysis, Writing – review \& editing. 
\textbf{Hiroyuki Ekawa:} Methodology, Software, Formal analysis, Investigation, Writing – review \& editing. \textbf{Samuel Escrig:} Methodology, Writing – review \& editing. 
\textbf{Yiming Gao:} Methodology, Writing – review \& editing. 
\textbf{Ayumi Kasagi:} Conceptualization, Methodology, Software, Validation, Formal analysis, Investigation, Data curation, Writing – review \& editing, Visualization. 
\textbf{Enqiang Liu:} Methodology, Software, Formal analysis, Investigation, Writing – review \& editing. 
\textbf{Abdul Muneem:} Methodology, Investigation, Writing – review \& editing. 
\textbf{Manami Nakagawa:} Conceptualization, Methodology, Software, Validation, Formal analysis, Investigation, Data curation, Writing – review \& editing.
\textbf{Kazuma Nakazawa:} Conceptualization, Methodology, Formal analysis, Investigation, Writing – review \& editing, Resources, Funding acquisition.
\textbf{Christophe Rappold:} Methodology, Formal analysis, Investigation, Writing – review \& editing.
\textbf{Nami Saito:} Conceptualization, Methodology, Software, Investigation, Writing – review \& editing. 
\textbf{Takehiko R. Saito:} Conceptualization, Methodology, Writing – original draft, Writing – review \& editing, Project administration, Supervision, Resources, Funding acquisition. 
\textbf{Shohei Sugimoto:} Methodology, Writing – review \& editing. 
\textbf{Masato Taki:} Conceptualization, Methodology, Software, Investigation, Writing – review \& editing.
\textbf{Yoshiki K. Tanaka:} Methodology, Investigation, Writing – review \& editing.
\textbf{He Wang:} Methodology, Investigation, Writing – review \& editing.
\textbf{Ayari Yanai:} Methodology, Writing – review \& editing.
\textbf{Junya Yoshida:} Conceptualization, Methodology, Software, Validation, Formal analysis, Investigation, Data curation, Writing – review \& editing. 
\textbf{Hongfei Zhang:} Methodology, Writing – review \& editing.

\section*{Declaration of competing interest}
The authors declare that they have no known competing financial interests or personal relationships that could have appeared to influence the work reported in this paper.

\section*{Data availability}
Data will be made available on request.

\section*{Acknowledgments}
This work was supported by JSPS KAKENHI Grant Numbers JP16H02180, JP20H00155, JP18H05403, and JP19H05147 (Grant-in-Aid for Scientific Research on Innovative Areas 6005). A.K was supported by JSPS KAKENHI Grant Numbers JP23K19051 (Grant-in-Aid for Research Activity Start-up).
The authors thank the J-PARC E07 collaboration for providing the emulsion sheets. The authors thank Risa Kobayashi, Michi Ando, Chiho Harisaki and Hanako Kubota of RIKEN and Yoko Tsuchii of Gifu University for their technical support in mining events in the J-PARC E07 nuclear emulsions. The authors thank Yukiko Kurakata of RIKEN including the administrative works.

% \bibliography{<your bibdatabase>}
\bibliographystyle{elsarticle-num}

\bibliography{elsarticle}





% \end{thebibliography}

\end{document}